%% file: main.tex
\documentclass[aps,superscriptaddress,nofootinbib]{revtex4-1}
\usepackage{graphicx}
\usepackage{color}
\graphicspath{{figures/}{fig/}}
\usepackage{amsmath}
\usepackage{amssymb}
\usepackage{bm}
\usepackage{slashed}
\usepackage{epsfig}
\usepackage{amsfonts}
\usepackage{epstopdf}
\usepackage{hyperref}
\usepackage{bbm}
\usepackage{textcomp}
\usepackage{color}

\begin{document}
\title{Electroweak loop corrections to $gg\to gH$ at the LHC}

\author{Huan-Yu Bi}
\email{bihy@hainanu.edu.cn}
\affiliation{School of Physics and
 Optoelectronic Engineering, Hainan University, Haikou, 570228, China}

\author{Yan-Qing Ma}
\email{yqma@pku.edu.cn}
\affiliation{School of Physics, Peking University, Beijing 100871, China}
\affiliation{Center for High Energy Physics, Peking University, Beijing 100871, China}

\author{Dao-Ming Mu}
\email{mudaoming@pku.edu.cn}
\affiliation{School of Physics, Peking University, Beijing 100871, China}

\date{\today}

\begin{abstract}
We present the results of the complete electroweak loop corrections to the process \( gg \to gH \) at the Large Hadron Collider. The electroweak corrections to the total cross section are found to be approximately \( +4\% \). At the differential level, the corrections exceed \( +4\% \) in the low Higgs transverse momentum region and fall below \( -4\% \) in the high transverse momentum region, leading to a noticeable shape distortion for this distribution. Our results represent a necessary step towards to complete next-to-leading order electroweak correction of the Higgs + jet process.
\end{abstract}
\maketitle
\allowdisplaybreaks
\section{Introduction}

As the cornerstone of the electroweak symmetry breaking mechanism, the Higgs boson plays a central role in the Standard Model (SM) of particle physics. Since its discovery by the ATLAS~\cite{ATLAS:2012yve} and CMS~\cite{CMS:2012qbp} collaborations at the Large Hadron Collider (LHC) in 2012, the precise study of Higgs-related processes has become a key avenue for testing the internal consistency of the SM, probing the dynamics of electroweak symmetry breaking, and searching for potential deviations that could signal new physics.

Among various Higgs production channels, the associated production of a Higgs boson with a jet—especially in the high-transverse-momentum regime—serves multiple important purposes. It provides a flexible channel for probing the strength of the Higgs self-coupling. It is also sensitive to potential modifications of the gluon–gluon–Higgs and Higgs-top yukawa interaction, which may indicate the presence of physics beyond the SM \cite{Schlaffer:2014osa}. Currently, the measurement of differential cross sections for Higgs + jet(s) production at the LHC exhibits statistical uncertainties at the level of 10\% to 15\% \cite{ATLAS:2022fnp,ATLAS:2020wny}. With the upcoming High-Luminosity LHC (HL-LHC), these uncertainties are expected to reduce by approximately a factor of 5 \cite{Huss:2025nlt}.

Theoretically, next-to-leading-order (NLO) QCD calculations incorporating the full top-quark mass dependence\cite{Jones:2018hbb,Chen:2021azt} as well as bottom-quark mass dependence\cite{Bonciani:2022jmb} have been completed. Moreover, in the heavy top-quark mass limit (HTL), next-to-next-to-leading-order (NNLO) QCD corrections have also been obtained \cite{Boughezal:2013uia,Chen:2014gva,Boughezal:2015dra,Boughezal:2015aha} and with partial top-quark mass effect by  supplementing the result by the full quark mass dependence at leading order (LO) \cite{Chen:2016zka}. Attempts are made for next-to-next-to-next-to-leading-order (NNNLO) QCD corrections in HTL, with results for many of  the relevant Feynman integrals \cite{Henn:2023vbd,Gehrmann:2023etk,Gehrmann:2024tds}.
At the same time, NLO electroweak (EW) corrections are also necessary to achieve the precision goal. While NLO electroweak corrections related to single Higgs production have been completed for the VBF \cite{Ciccolini:2007jr,Denner:2014cla}, $VH$\cite{Ciccolini:2003jy,Denner:2011id,Obul:2018psx,Granata:2017iod}, $VHj$ \cite{Granata:2017iod}, and $t\bar{t}H$ \cite{Zhang:2014gcy,Frixione:2014qaa,Denner:2016wet} channels, a comprehensive treatment for the $gg\to Hg$ channel remains unavailable, due to the involvement of complex two-loop Feynman integrals with multiple mass scales. The NLO EW corrections for the $gg\to Hg$ channel have so far only been partially studied.
In Refs.~\cite{Bonetti:2020hqh, Becchetti:2020wof}, the authors computed the electroweak corrections under the approximation of including only light-fermion loop contributions. While in Ref.~\cite{Davies:2023npk}, the electroweak corrections were investigated in the large-$m_t$ limit.  Refs.~\cite{Gao:2023bll, Haisch:2024nzv} studied the impact of diagrams involving the Higgs boson trilinear self-coupling.

In this paper, we present the first computation of  EW loop corrections to the process \( gg \to H g \), including the full top-quark mass dependence. The most challenging aspect of this calculation lies in the evaluation of the two-loop Feynman integrals, especially the so-called master integrals (MIs). These MIs are computed numerically using the differential equation method \cite{Kotikov:1990kg,Remiddi:1997ny,Caffo:2008aw,Czakon:2008zk}, with boundary conditions provided by auxiliary mass flow method \cite{Liu:2017jxz,Liu:2021wks,Liu:2022mfb,Liu:2022chg}. To facilitate efficient and accurate evaluations across the entire phase space, we construct a series of asymptotic expansions of squared amplitude for one dimensional interpolating. 
This work represents a necessary step forward in completing the theoretical description of the \( H+\text{jet} \) process at high precision.

The paper is organized as follows. In section \ref{sec2}, we present the detail of the calculation. In section \ref{sec3}, we summarize the computational setup
 of the calculations employed in our study. In section \ref{sec4}, we show our phenomenological results including total and  differential cross sections. Finally, we give our conclusions in section \ref{sec5}.

\section{Calculation strategies}\label{sec2}

Typically, for NLO EW corrections, the loop (or virtual) corrections need to be combined with the photon radiation process to obtain a finite result. However, in the context of the current problem, a finite result can be obtained by considering only the loop corrections to the lowest-order topology  (one loop) of $gg \to Hg$. The reason is that neither the external gluons nor the Higgs boson can emit photons directly, while photon emission from internal legs cannot generate infrared or collinear divergences in this process. We thus compute EW loop corrections to $gg \to Hg$ in this work, serving as a necessary step towards the full  NLO EW corrections to $H+$jet process. 

\begin{figure}[htbp]
    \includegraphics[width=1\textwidth]{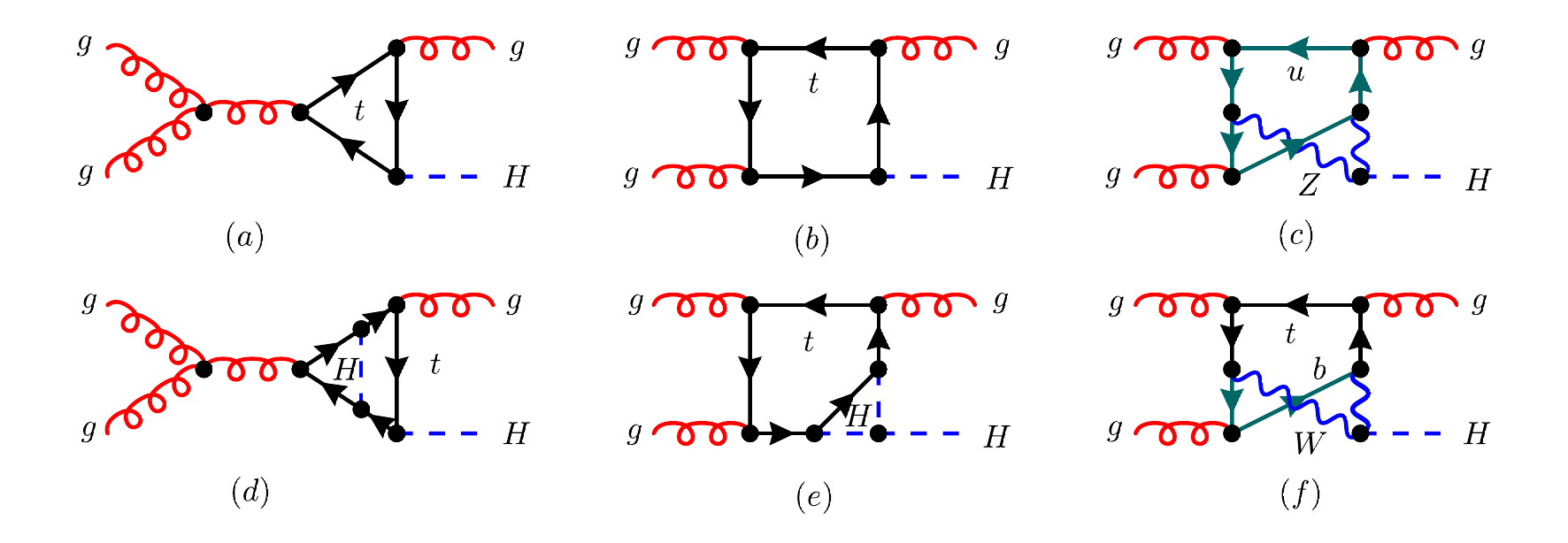}
\caption{\label{fig:feynmandiagrams}
Representative Feynman diagrams for $gg\to gH$ at LO ($a$)-($b$) and EW loop corrections ($c$)-($f$). }
\end{figure}

We generated the Feynman diagrams and amplitudes for this process using {\tt FeynArts} \cite{Hahn:2000kx}, and representative Feynman diagrams are shown in Fig.~\ref{fig:feynmandiagrams}. As the $\mathcal{S}$-operator is a color-singlet operator, the Wigner-Eckart theorem implies that the amplitude for the process $g_{a_1}(p_1)\,g_{a_2}(p_2)\to g_{a_3}(p_3)\,H(p_4)$ takes the form
\begin{equation}
    M^{a_1 a_2 a_3} = \langle p_1,a_1;\,p_2,a_2|\mathcal{S}|p_3,a_3;\,p_4 \rangle 
    = \langle a_1,a_2|a_3 \rangle \, \langle p_1,p_2|\mathcal{S}|p_3,p_4 \rangle 
    = f^{a_1 a_2 a_3} \, \langle p_1,p_2|\mathcal{S}|p_3,p_4 \rangle,
\end{equation}
where $a_i\ (i=1,2,3)$ represent the color indices of the gluons and $f^{a_1a_2a_3}$ is the QCD structure constant.
Hence, the scattering amplitude can thus be expressed as
\begin{equation}
    M^{a_1a_2a_3}(p_{1},p_{2},p_{3})=f^{a_{1} a_{2} a_{3}} \epsilon_{1}^{\mu} \epsilon_{2}^{\nu} (\epsilon_{3}^{\rho})^* \mathcal{M}_{\mu \nu \rho}\left(\hat{s}, \hat{t}, \hat{u} \right), 
\end{equation}
where  $\epsilon_i\ (i=1,2,3)$ are the gluon polarization vectors, and the Mandelstam variables are 
\begin{equation}
\hat{s}=\left(p_{1}+p_{2}\right)^{2}, \quad \hat{t}=\left(p_{1}-p_{3}\right)^{2}, \quad \hat{u}=\left(p_{2}-p_{3}\right)^{2} .
\end{equation}
In general, the amplitude can be decomposed into the following tensor structures
\begin{equation}
     \mathcal{M}^ {\mu \nu \rho }  (\hat{s},\hat{t},\hat{u})=  F_ {1}   g^ {\mu \nu }   p_ {2}^ {\rho }  +  F_ {2}   g^ {\mu \rho }   p_ {1}^ {\nu }  +  F_ {3}   g^ {\nu  \rho} p_ {3}^{\mu}  +  F_ {4}   p_ {3}^ {\mu }   p_ {1}^{\nu}   p_ {2}^ {\rho } + \Delta_5^{\mu \nu \rho}+\Delta_0^{\mu \nu \rho}, 
\end{equation}
where $F_i(i=1,2,3,4)$ are the so called form factors. $\Delta_5^{\mu\nu\rho}$ is linearly dependent on the Levi-Civita tensor, and arises for the first time at EW loop corrections.  Owing to the fact that there are only three independent external momenta, any contraction of $\Delta_5^{\mu\nu\rho}$ with external momenta yields zero. $\Delta_0^{\mu\nu\rho}$ contains all other tensor structures. Following the conventions introduced in Ref.~\cite{Melnikov:2016qoc}, using the transversality conditions and choosing a cyclic gauge fixing condition 
\begin{equation}
    \epsilon_i\cdot p_i=0,\quad \epsilon_1\cdot p_2=\epsilon_2\cdot p_3=\epsilon_3\cdot p_1=0,
\end{equation}
$\Delta_0^{\mu\nu\rho}$ does not contribute to $M$. Thus, only four form factors contribute to the EW loop corrections, and despite the appearance of $\Delta_5^{\mu\nu\rho}$, the projectors defined in Ref.~\cite{Melnikov:2016qoc} remain valid for extracting the form factors from $\mathcal{M}$.

At two-loop level, when there is a single $\gamma_5$ in the fermion loop(s),  the resulting $\Delta_5^{\mu\nu\rho}$ does not contribute to the corrections, as discussed previously. In cases where there are two $\gamma_5$ in two different fermion loops, such contribution vanishes due to the color algebra. The only non-vanishing case involving $\gamma_5$ is the one where two $\gamma_5$ matrices appear in the same fermion loop. Similarly to Ref.~\cite{Bi:2023bnq}, we adopt the simplest naive $\gamma_5$ scheme by directly applying the anticommutation relation $\left\{ \gamma_5,\gamma_{\mu}\right\}=0 $.

Using the {\tt {CalcLoop}} package \cite{calcloop}, the form factors are expressed as linear combinations of scalar Feynman integrals, which are categorized into 3 (108) integral families based on the type of propagators at the one-loop (two-loop) level. Subsequently, we use {\tt Blade} \cite{Guan:2024byi} combining with {\tt FiniteFlow} \cite{Peraro:2019svx} to reduce the loop integrals in each family to a simpler set of MIs. Since the final finite physical result is insensitive to the small dimensional regulator $\epsilon$ with $D = 4 - 2\epsilon$, we set $\epsilon = 1/1000$ throughout the entire calculation \cite{Liu:2022mfb, Liu:2022chg}. This strategy allows us to eliminate one variable during the calculation and avoid repeated computations of MIs with different  values of $\epsilon$ to obtain a Laurent expansion. This is expected to introduce an $\mathcal{O}(\epsilon)$ error in the final result, which can be further reduced to $\mathcal{O}(\epsilon^2)$ by combining the result with that obtained using $\epsilon = -1/1000$.

\begin{figure}[htbp]
    \includegraphics[width=0.4\textwidth, trim = 12cm 7.3cm 10cm 6.5cm, clip]{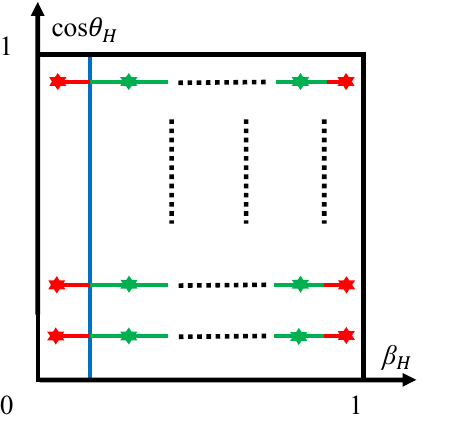}
\caption{\label{grid}
Vertical line denotes the asymptotic expansions in $\cos\theta_H$, which provide boundary conditions for solving the differential equations in $\beta_H$. The colored horizontal lines represent the asymptotic expansions in $\beta_H$ at different expansion points (indicated by stars), evaluated at ten selected $\cos\theta_H$ values.}
\end{figure}

The most time consuming part in the calculation is the computing of the MIs. To speed up the calculation, we construct asymptotic expansions in variable $\beta_H=({\hat{s}-m_H^2})/({\hat{s}+m_H^2})$ at several fixed values of $\cos{\theta_H}=({\hat{u}-\hat{t}})/({\hat{s}-m_H^2})$, as shown in red and green lines in Fig.~\ref{grid}, such that one-dimensional interpolation along the $\cos\theta_H$ direction becomes feasible. Since the squared amplitude is an even function of $\cos\theta_H$, it suffices to consider only $\cos\theta_H \geq 0$.

We use {\tt Blade} to construct the differential equations analytically in each family with respect to the variables $\beta_H$ and $\cos{\theta_H}$. With  the boundary condition provided by {\tt AMFlow} package \cite{Liu:2022chg} implementing the auxiliary mass flow method \cite{Liu:2017jxz,Liu:2022mfb,Liu:2021wks}, we solve the differential equations in $\cos\theta_H$ at a fixed value $\beta_H = {28}/{93}$.  We then construct asymptotic expansions valid for $\cos\theta_H \in [0,1]$, corresponding to the blue line in Fig.~\ref{grid}.
The asymptotic expansion in $\cos\theta_H$ at $\beta_H = {28}/{93}$ serves as the boundary conditions for solving the differential equations in $\beta_H$, resulting in asymptotic expansions respect to $\beta_H$, shown as red and green lines in Fig.~\ref{grid}. In this work, the asymptotic expansions for $\beta_H$  are evaluated in the range $\beta_H \in [1/5,9999/10000]$ corresponding to $p_T^H \geq 25.5$ Gev and $\sqrt{s} \leq  17.6$ TeV, which is sufficient for the phenomenology studies at the LHC. When solving the differential equations with $\beta_H$, singularities in the MIs arise from internal particles going on shell, corresponding to $\sqrt{\hat{s}} = m_W$, $m_Z$, $2m_t$, $m_t + m_W$, $2m_t + m_Z$, or $2m_t + m_H$ in the physical region. Analytical continuation is performed by introducing an infinitesimal positive imaginary part to $\hat{s}$ across these thresholds.

The asymptotic expansion of MIs in $\beta_H$ are computed at 10 chosen values of $\cos\theta_H = \{\frac{1}{4},\frac{101}{200},\frac{5}{8},\frac{3}{4},\frac{7}{8},$ $\frac{15}{16},$ $\frac{31}{32}$, $\frac{63}{64}$, $\frac{127}{128},\frac{255}{256}\}$. Then we obtain the asymptotic expansion of the squared amplitude for LO and EW loop corrections, denoted as $|M_{\rm LO}|^2$ and $2{\rm Re}(M_{\rm LO}*M_{\rm NLO}^*)$, respectively. For any desired value of $\cos\theta_H$, $|M_{\rm LO}|^2$ and $2{\rm Re}(M_{\rm LO}\times M_{\rm NLO}^*)$ are obtained via a cubic polynomial interpolation using the four nearest points chosen from the set of previously computed values of $\cos\theta_H$ and their negatives. As a consequence, the $\mathcal{K}$-factor for the squared amplitude can be computed at any phase space point in a very efficient way.

\begin{figure}[htbp]
    \includegraphics[width=0.6\textwidth]{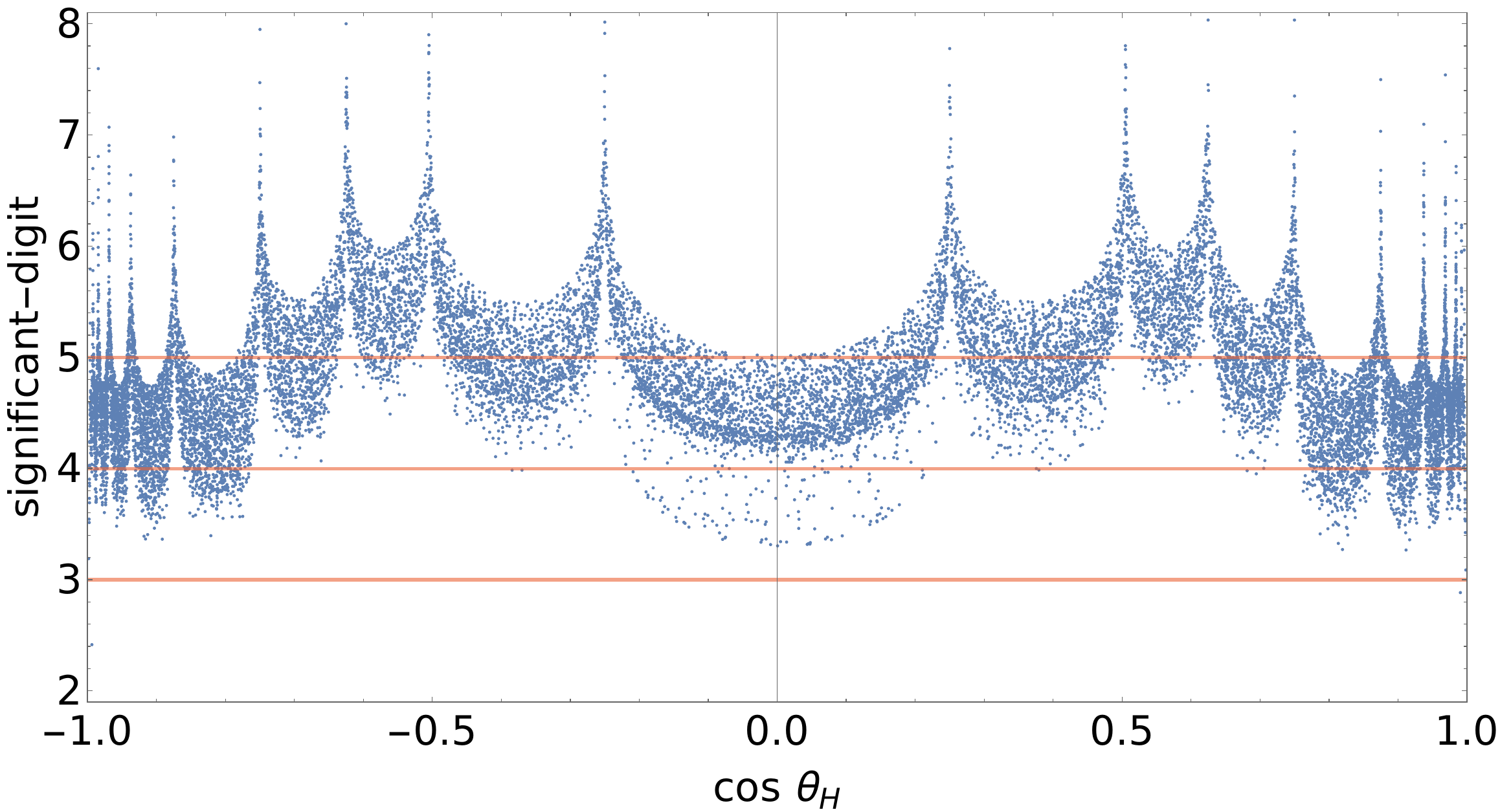}
\caption{\label{fig:cosdigits} Significant-digit agreement between interpolated and directly computed $\mathcal{K}$-factors over $2 \times 10^{4}$ sampled phase-space points.}
\end{figure}

To validate the accuracy of the interpolation,  we computed the  MIs at $2 \times 10^{4}$ different phase space points generated by importance sampling. These MIs are calculated by solving the differential equations numerically with sufficient high precision. Then the $\mathcal{K}$-factor for the squared amplitude are computed and compared to the $\mathcal{K}$-factor calculated based on the interpolation. Fig.~\ref{fig:cosdigits} illustrates the interpolation accuracy across the sampled phase-space points. We find 3-digit agreement for 99.99\% of the phase space points, 4-digit agreement for 94\% of the phase space points, and 5-digit agreement for 37\% of the phase space points. The worst agreement corresponds to 2.4 significant digits, occurring at the point $(\beta_H, \cos\theta_H) \approx (0.9998, -0.99)$. In this region, where both $\beta_H$ and $|\cos\theta_H|$ are close to 1, the $\mathcal{K}$-factor of squared amplitude exhibits rapid variations with respect to $\cos\theta_H$, which explains the reduced interpolation accuracy. The two different methods lead to a relative difference of $2.4 \times 10^{-5}$ for NLO cross sections based the $2 \times 10^4$ sample points, which ensures the correctness of the interpolation algorithm. 

The phase space integration is carried out and optimized using {\tt{Parni}} \cite{vanHameren:2007pt}. A total of $1\times10^8$ events are generated at the LO, and the results are cross-checked with {\tt{MadGraph5}} \cite{Alwall:2014hca}. The one-loop Feynman integrals in the LO calculation are carried out by {\tt DCT} package \cite{Huang:2024qan}, which enables efficient evaluation of one-loop integrals to arbitrary orders in $\epsilon$. LO events are then stored in the form of modified Les Houches Event Files \cite{Alwall:2006yp}, enabling reweighting to the NLO events.

\section{Input parameters}\label{sec3}
We adopt the following SM parameters
\begin{eqnarray}
\frac{m_H^2}{m_t^2}=\frac{12}{23},~\frac{m_Z^2}{m_t^2}=\frac{23}{83},~\frac{m_W^2}{m_t^2}=\frac{14}{65},
\end{eqnarray}
 which correspond to the numerical values \( m_H \approx 124.74~\text{GeV} \), \( m_Z \approx 90.91~\text{GeV} \), and \( m_W \approx 80.14~\text{GeV} \), taking the top quark mass as \( m_t = 172.69~\text{GeV} \)~\cite{ParticleDataGroup:2022pth}. All other masses and widths of particles are set to zero.  The relevant masses and fields are  renormalized in the on-shell scheme, while the renormalization of  $\alpha$ is conducted in the $G_{\mu}$ scheme \cite{Denner:2019vbn}.
In this scheme, the fine structure constant is given by 
\begin{eqnarray}
\alpha = \alpha_{G_\mu} = \frac{\sqrt{2}}{\pi} G_\mu m_W^2 \left(1 - {m_W^2}/{m_Z^2} \right) \approx {1}/{133.12},
\end{eqnarray}
with \( G_\mu = 1.166378 \times 10^{-5}~\text{GeV}^{-2} \).  This fine structure constant scheme is suitable for EW corrections when there are large EW Sudakov logarithms caused by the soft or collinear weak gauge boson exchange at high energies \cite{Denner:2014bna,Denner:2019vbn}. 
We neglect quark mixing among different generations by assuming a diagonal Cabibbo–Kobayashi–Maskawa (CKM) matrix.
For both LO and EW loop corrections, we employ the \texttt{NNPDF31\_nlo\_as\_0118} PDF set~\cite{NNPDF:2017mvq}. The running of the strong coupling constant \( \alpha_s \) is performed at two-loop accuracy using the \texttt{LHAPDF6} library~\cite{Buckley:2014ana}, assuming five active quark flavors.
The central value of renormalization and factorization scales are set dynamically as the sum of transverse masses divided by two for all final-state particles:
\begin{eqnarray}
\mu_0 = \frac{1}{2} \sqrt{p_T^2 + m_H^2} + \frac{1}{2} p_T,
\end{eqnarray}
where $p_T$ is the magnitude of transverse momentum of the final-sate gluon, equaling to the magnitude of transverse momentum of the Higgs. We impose a kinematic constraint \( p_T \geq 30~\text{GeV} \) in the numerical study.
To estimate theoretical uncertainty, we vary \( \mu_r \) and \( \mu_f \) independently around the central scale \( \mu_0 \) within the range
\begin{eqnarray}
\frac{1}{2} \leq \frac{\mu_r}{\mu_0}, \frac{\mu_f}{\mu_0} \leq 2, \quad \text{with} \quad \frac{1}{2} \leq \frac{\mu_r}{\mu_f} \leq 2,
\end{eqnarray}
resulting in the following seven-point scale variations:
\begin{eqnarray}
\left(\frac{\mu_r}{\mu_0}, \frac{\mu_f}{\mu_0}\right) = \{ (2,1), (0.5,1), (1,2), (1,1), (1,0.5), (2,2), (0.5,0.5) \}.
\end{eqnarray}

\section{Numerical results}\label{sec4}
The LO and EW loop correction results are presented in Tab.~\ref{total-cs}, with $\sqrt{s} = 13$ TeV. The results exhibit big renormalization scale uncertainties, amounting to approximately 39\% for both LO and EW loop corrections when the factorization scale is fixed. In contrast, the factorization scale dependence is moderate, around 2\% for both LO and EW loop corrections when the renormalization scale is fixed. 
Taking into account the full 7-point scale variations, we obtain
\begin{eqnarray}
\sigma_{\rm LO}=6.37^{+2.54(40\%)}_{-1.71(27\%)},\\
\sigma_{\rm EW}=6.63^{+2.64(40\%)}_{-1.78(27\%)}.
\end{eqnarray}
On the other hand, the ${\cal K}$-factor, which is defined as ${\cal K}=\sigma_{\rm EW}/\sigma_{\rm LO}$, remains remarkably stable at ${\cal K} = 1.041$ across different scale choices, similar to double Higgs production in Ref.~\cite{Bi:2023bnq}. Therefore, it is sufficient for this work to compute the ${\cal K}$-factor with $\mu_f=\mu_r=\mu_0$, as the scale dependence can be further reduced by including higher-order QCD corrections~\cite{Jones:2018hbb,Chen:2021azt,Bonciani:2022jmb,Boughezal:2013uia,Chen:2014gva,Boughezal:2015dra,Boughezal:2015aha,Chen:2016zka}.
\begin{table}[htbp]
\begin{center}
\setlength{\tabcolsep}{1.3pt} 
\renewcommand\arraystretch{1.3}
\begin{tabular}{cccccccc}
\hline
\hline
$(\mu_r,\mu_f)$ & $(\mu_0,\mu_0)$  & $(\mu_0,0.5\mu_0)$  & $(\mu_0,2\mu_0)$ & $(0.5\mu_0,\mu_0)$ &$(2\mu_0,\mu_0)$&$(0.5\mu_0,0.5\mu_0)$&$(2\mu_0,2\mu_0)$ \\
\hline
$\sigma_{\rm LO}$   & 6.37(2)   & 6.43(2)  & 6.25(2) & 8.83(2) & 4.75(1) & 8.91(2) & 4.66(1)\\
$\sigma_{\rm EW}$  & 6.63(2)   & 6.69(2)  & 6.51(2) &  9.20(2) & 4.95(1) & 9.27(2) & 4.85(1)\\
${\cal K}$-factor  & 1.041   &    1.041   &   1.041&1.041 & 1.041 & 1.041 & 1.041     \\
\hline
\hline
\end{tabular}
\caption{LO and EW loop corrected integrated cross sections (in pb) for $gg\to gH$ at the LHC with $\sqrt{s}= 13$ TeV. The numbers in the parentheses represent the statistical errors in phase space integration.}
\label{total-cs}
\end{center}
\end{table}

\begin{figure}[htbp]
    \includegraphics[width=0.5\textwidth]{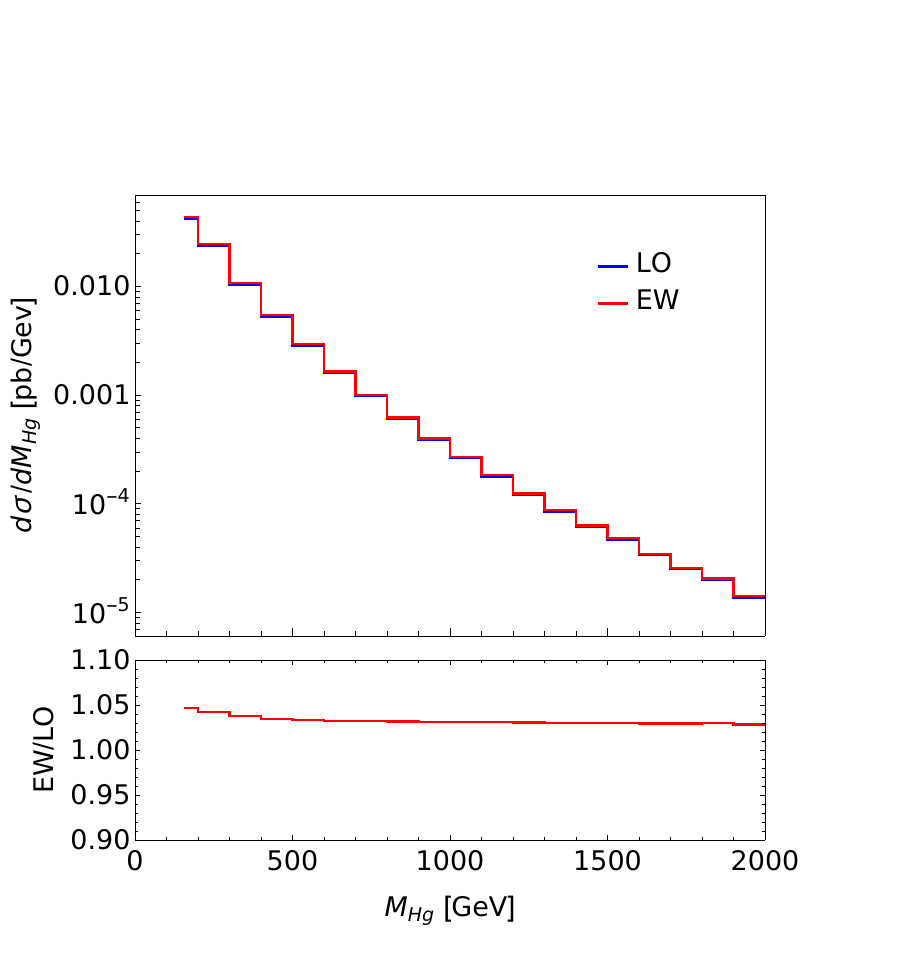}
\caption{\label{mhg}
Invariant mass distribution of the $Hg$ system  with $\sqrt{s}= 13$ TeV. The upper plot shows absolute predictions, and the lower panel displays the differential ${\cal K}$-factor.}
\end{figure}

\begin{figure}[htbp]
    \includegraphics[width=0.5\textwidth]{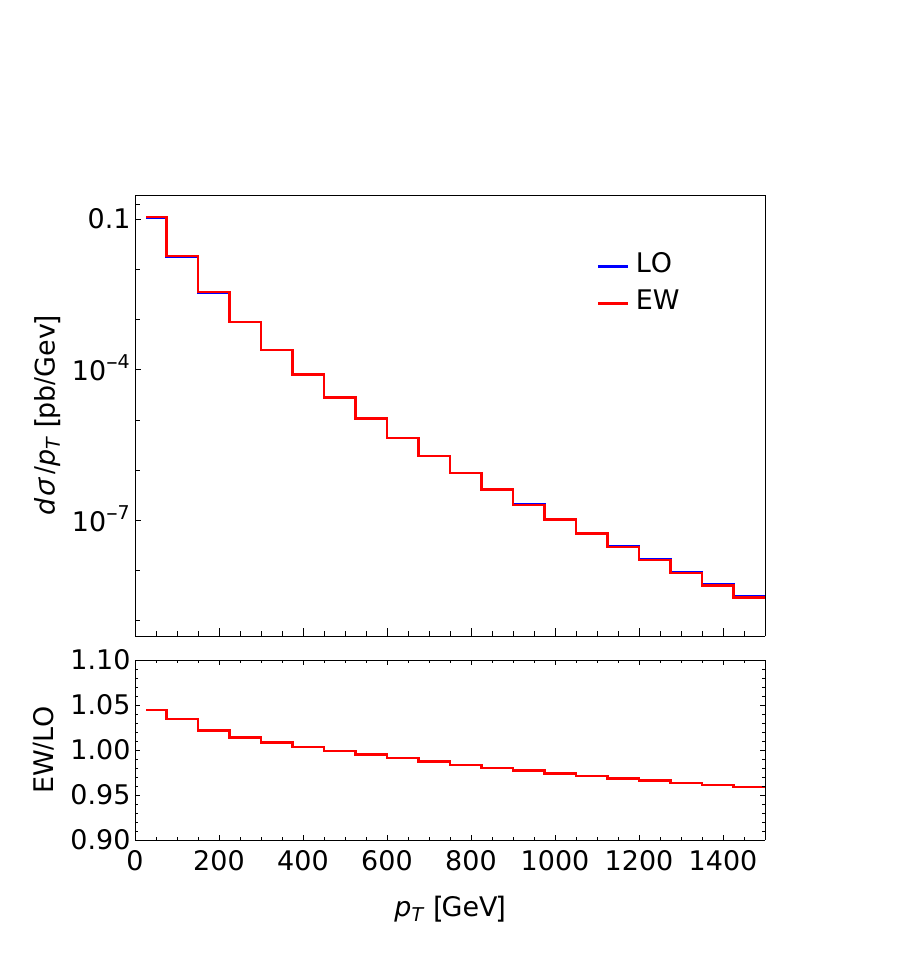}
\caption{\label{pth}
Transverse momentum distribution of  the  Higgs boson  with $\sqrt{s}= 13$ TeV. The upper plot shows absolute predictions, and the lower panel displays the differential ${\cal K}$-factor.}
\end{figure}

\begin{figure}[htbp]
    \includegraphics[width=0.5\textwidth]{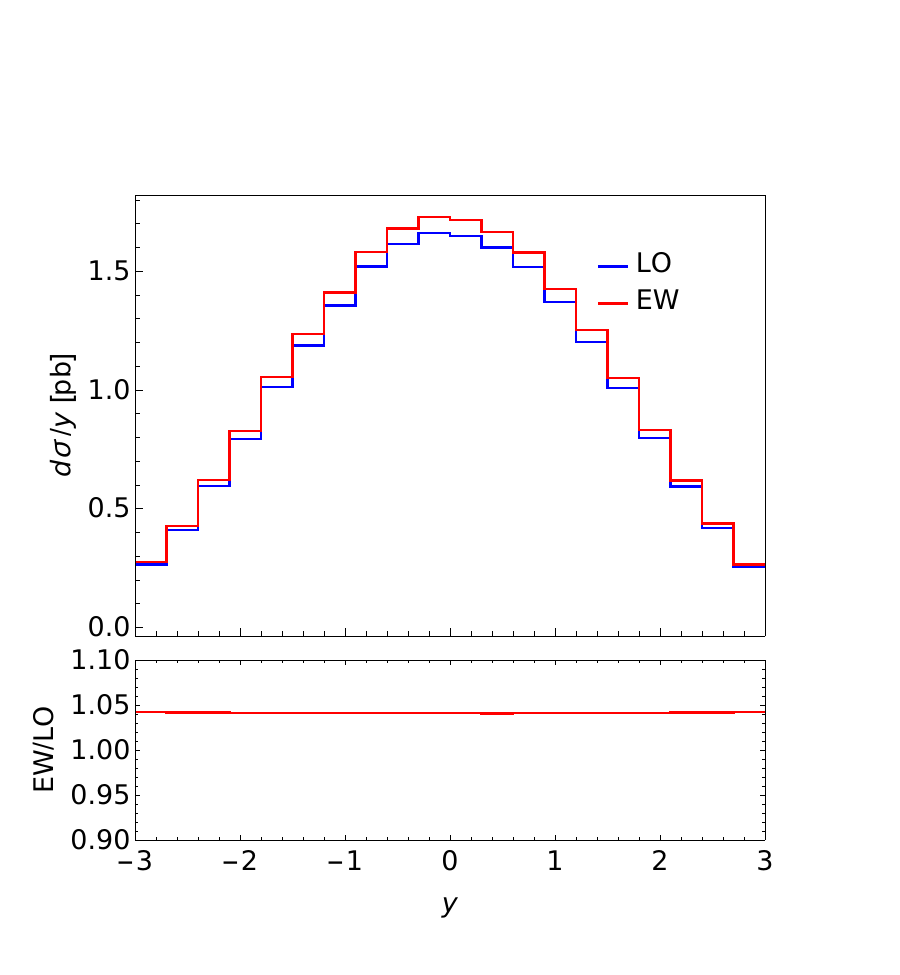}
\caption{\label{yh}
Rapidity distribution of  the  Higgs boson  with $\sqrt{s}= 13$ TeV. The upper plot shows absolute predictions, and the lower panel displays the differential ${\cal K}$-factor.}
\end{figure}

Differential cross sections provide rich insights into the underlying physics, whether within the SM or in extensions beyond it. Notably, EW loop corrections can have a non-uniform effect, showing stronger or weaker impacts in certain regions of phase space compared to their influence on the total cross section. Since the main concern of this paper is the ${\cal K}$-factor of the EW corrections,  in the following we only present the differential cross section based on the central scale $\mu_0$.
In Fig.~\ref{mhg}, we present the invariant mass distribution of the $Hg$ system. The plot indicates that the ${\cal K}$-factor decreases fast at the beginning of the spectrum and then decreases slowly at high energy region. The curve in the lower panel stays slightly above 1, around 1.03 to 1.05, showing a modest positive EW loop correction across the entire $M_{Hg}$ range we have shown. In Fig.~\ref{pth}, we present the transverse momentum distribution of the Higgs boson. 
The curves in the upper panel decrease steeply with increasing $p_T$, indicating that Higgs bosons are predominantly produced with low $p_T$. However, as discussed in the introduction, the region of high $p_T$ Higgs boson is of particular interest. In the lower panel of Fig.~\ref{pth}, we show the differential ${\cal K}$-factor. Similar to the behavior observed in the $M_{Hg}$ distribution, the ${\cal K}$-factor for $p_T$ decreases  rapidly at the beginning of the spectrum. However, in contrast to the $M_{Hg}$ cases, the ${\cal K}$-factor becomes negative around $p_T=400$ GeV and continues to decrease in the  high $p_T$ region. Consequently, EW corrections become increasingly significant with rising collision energy. Finally we present the Higgs rapidity distribution in Fig.~\ref{yh}, a nearly flat ${\cal K}$-factor is observed, approximately 1.04, similar to the total cross section.

\section{Summary}\label{sec5}
We present a detailed analysis of the EW loop corrections to the process $gg\to Hg$ at the LHC with $\sqrt{s}=13$ TeV. In the computation, the involved Feynman loop integrals are evaluated numerically with the differential equation methods. Their asymptotic expansions are carried out with respect to the variable $\beta_H$ and the expansion points are carefully chosen to account for the singularities in $\beta_{H}-{\rm cos}\theta_{H}$ plane. This ensures that the asymptotic expansion covers the region of $\beta \in [1/5,9999/10000]$ which is valid for $p_T \geq 25.5$ Gev and $\sqrt{s} \leq  17.6$ TeV. Based on these asymptotic expansions, the squared matrix elements at both LO and EW loop levels are expressed in a compact and efficient asymptotic expansion form. Using the asymptotic expansions, the angular dependence is evaluated at 10 different ${\rm cos}\theta_H$ values, allowing the evaluation at arbitrary scattering angles via one-dimensional interpolation. This is more efficient and accurate than conventional grid-based methods, which require two-dimensional interpolation. We plan to make our asymptotic expressions for the squared matrix elements publicly available in the near future, enabling straightforward integration with event generators.

Our  results shows that the EW corrections to the total cross section is approximately $4.1\%$, given the applied  $p_T$ cutoff and chosen input parameters. While the scale uncertainties in the total cross section are significant, the $\cal K$-factor due to EW corrections  remains remarkably stable under scale variations. 
For the differential cross sections, particularly  the $p_T$ distribution, the EW corrections exhibit a shape-dependent behavior: 
they are about +4\% at the low $p_T$ and decrease to -4\% at at high $p_T$. This highlight the importance of EW corrections in precision studies of Higgs production in the high $p_T$ region at the LHC.
\begin{acknowledgments}
We thank Xin Guan and Rui-Jun Huang for many helpful discussions. The work was supported in part by the National Natural Science Foundation of China (Grants No. 12325503) and the High-performance Computing Platform of Peking University. Feynman diagrams are drawn using the {\tt FeynGame} program \cite{Harlander:2020cyh}.
\end{acknowledgments}

{\it{Note added:}} While our paper is being finalized, a preprint by Chen et al. \cite{Chen:2025sgo} appeared, in which these authors reached similar conclusion as ours.


\input{ref.bbl}
\end{document}

%% file: ref.bbl
\providecommand{\href}[2]{#2}\begingroup\raggedright\endgroup

%% file: main.bbl
\begin{thebibliography}{10}

\bibitem{ATLAS:2012yve}
{\bfseries ATLAS} , G.~Aad {\em et al.}, {\it {Observation of a new particle in
  the search for the Standard Model Higgs boson with the ATLAS detector at the
  LHC}},  \href{http://dx.doi.org/10.1016/j.physletb.2012.08.020}{{\em Phys.
  Lett. B} {\bfseries 716} (2012) 1--29}
  [\href{http://arxiv.org/abs/1207.7214}{{\ttfamily arXiv:1207.7214}}]
  [\href{http://inspirehep.net/search?p=find+ATLAS:2012yve}{{\ttfamily
  InSPIRE}}].

\bibitem{CMS:2012qbp}
{\bfseries CMS} , S.~Chatrchyan {\em et al.}, {\it {Observation of a New Boson
  at a Mass of 125 GeV with the CMS Experiment at the LHC}},
  \href{http://dx.doi.org/10.1016/j.physletb.2012.08.021}{{\em Phys. Lett. B}
  {\bfseries 716} (2012) 30--61}
  [\href{http://arxiv.org/abs/1207.7235}{{\ttfamily arXiv:1207.7235}}]
  [\href{http://inspirehep.net/search?p=find+CMS:2012qbp}{{\ttfamily
  InSPIRE}}].

\bibitem{Schlaffer:2014osa}
M.~Schlaffer, M.~Spannowsky, M.~Takeuchi, A.~Weiler, and C.~Wymant, {\it
  {Boosted Higgs Shapes}},
  \href{http://dx.doi.org/10.1140/epjc/s10052-014-3120-z}{{\em Eur. Phys. J. C}
  {\bfseries 74} (2014) 3120} [\href{http://arxiv.org/abs/1405.4295}{{\ttfamily
  arXiv:1405.4295}}]
  [\href{http://inspirehep.net/search?p=find+Schlaffer:2014osa}{{\ttfamily
  InSPIRE}}].

\bibitem{ATLAS:2022fnp}
{\bfseries ATLAS} , G.~Aad {\em et al.}, {\it {Measurements of the Higgs boson
  inclusive and differential fiducial cross-sections in the diphoton decay
  channel with pp collisions at $ \sqrt{s} $ = 13 TeV with the ATLAS
  detector}},  \href{http://dx.doi.org/10.1007/JHEP08(2022)027}{{\em JHEP}
  {\bfseries 08} (2022) 027} [\href{http://arxiv.org/abs/2202.00487}{{\ttfamily
  arXiv:2202.00487}}]
  [\href{http://inspirehep.net/search?p=find+ATLAS:2022fnp}{{\ttfamily
  InSPIRE}}].

\bibitem{ATLAS:2020wny}
{\bfseries ATLAS} , G.~Aad {\em et al.}, {\it {Measurements of the Higgs boson
  inclusive and differential fiducial cross sections in the 4$\ell $ decay
  channel at $\sqrt{s}$ = 13 TeV}},
  \href{http://dx.doi.org/10.1140/epjc/s10052-020-8223-0}{{\em Eur. Phys. J. C}
  {\bfseries 80} (2020) 942} [\href{http://arxiv.org/abs/2004.03969}{{\ttfamily
  arXiv:2004.03969}}]
  [\href{http://inspirehep.net/search?p=find+ATLAS:2020wny}{{\ttfamily
  InSPIRE}}].

\bibitem{Huss:2025nlt}
A.~Huss, J.~Huston, S.~Jones, M.~Pellen, and R.~R{\"o}ntsch, {\it {Les Houches
  2023 -- Physics at TeV Colliders: Report on the Standard Model Precision
  Wishlist}},  [\href{http://arxiv.org/abs/2504.06689}{{\ttfamily
  arXiv:2504.06689}}]
  [\href{http://inspirehep.net/search?p=find+Huss:2025nlt}{{\ttfamily
  InSPIRE}}].

\bibitem{Jones:2018hbb}
S.~P. Jones, M.~Kerner, and G.~Luisoni, {\it {Next-to-Leading-Order QCD
  Corrections to Higgs Boson Plus Jet Production with Full Top-Quark Mass
  Dependence}},  \href{http://dx.doi.org/10.1103/PhysRevLett.120.162001}{{\em
  Phys. Rev. Lett.} {\bfseries 120} (2018) 162001}
  [\href{http://arxiv.org/abs/1802.00349}{{\ttfamily arXiv:1802.00349}}]
  [\href{http://inspirehep.net/search?p=find+Jones:2018hbb}{{\ttfamily
  InSPIRE}}]. [Erratum: Phys.Rev.Lett. 128, 059901 (2022)].

\bibitem{Chen:2021azt}
X.~Chen, A.~Huss, S.~P. Jones, M.~Kerner, J.~N. Lang, J.~M. Lindert, and
  H.~Zhang, {\it {Top-quark mass effects in H+jet and H+2 jets production}},
  \href{http://dx.doi.org/10.1007/JHEP03(2022)096}{{\em JHEP} {\bfseries 03}
  (2022) 096} [\href{http://arxiv.org/abs/2110.06953}{{\ttfamily
  arXiv:2110.06953}}]
  [\href{http://inspirehep.net/search?p=find+Chen:2021azt}{{\ttfamily
  InSPIRE}}].

\bibitem{Bonciani:2022jmb}
R.~Bonciani, V.~Del~Duca, H.~Frellesvig, M.~Hidding, V.~Hirschi, F.~Moriello,
  G.~Salvatori, G.~Somogyi, and F.~Tramontano, {\it {Next-to-leading-order QCD
  corrections to Higgs production in association with a jet}},
  \href{http://dx.doi.org/10.1016/j.physletb.2023.137995}{{\em Phys. Lett. B}
  {\bfseries 843} (2023) 137995}
  [\href{http://arxiv.org/abs/2206.10490}{{\ttfamily arXiv:2206.10490}}]
  [\href{http://inspirehep.net/search?p=find+Bonciani:2022jmb}{{\ttfamily
  InSPIRE}}].

\bibitem{Boughezal:2013uia}
R.~Boughezal, F.~Caola, K.~Melnikov, F.~Petriello, and M.~Schulze, {\it {Higgs
  boson production in association with a jet at next-to-next-to-leading order
  in perturbative QCD}},  \href{http://dx.doi.org/10.1007/JHEP06(2013)072}{{\em
  JHEP} {\bfseries 06} (2013) 072}
  [\href{http://arxiv.org/abs/1302.6216}{{\ttfamily arXiv:1302.6216}}]
  [\href{http://inspirehep.net/search?p=find+Boughezal:2013uia}{{\ttfamily
  InSPIRE}}].

\bibitem{Chen:2014gva}
X.~Chen, T.~Gehrmann, E.~W.~N. Glover, and M.~Jaquier, {\it {Precise QCD
  predictions for the production of Higgs + jet final states}},
  \href{http://dx.doi.org/10.1016/j.physletb.2014.11.021}{{\em Phys. Lett. B}
  {\bfseries 740} (2015) 147--150}
  [\href{http://arxiv.org/abs/1408.5325}{{\ttfamily arXiv:1408.5325}}]
  [\href{http://inspirehep.net/search?p=find+Chen:2014gva}{{\ttfamily
  InSPIRE}}].

\bibitem{Boughezal:2015dra}
R.~Boughezal, F.~Caola, K.~Melnikov, F.~Petriello, and M.~Schulze, {\it {Higgs
  boson production in association with a jet at next-to-next-to-leading
  order}},  \href{http://dx.doi.org/10.1103/PhysRevLett.115.082003}{{\em Phys.
  Rev. Lett.} {\bfseries 115} (2015) 082003}
  [\href{http://arxiv.org/abs/1504.07922}{{\ttfamily arXiv:1504.07922}}]
  [\href{http://inspirehep.net/search?p=find+Boughezal:2015dra}{{\ttfamily
  InSPIRE}}].

\bibitem{Boughezal:2015aha}
R.~Boughezal, C.~Focke, W.~Giele, X.~Liu, and F.~Petriello, {\it {Higgs boson
  production in association with a jet at NNLO using jettiness subtraction}},
  \href{http://dx.doi.org/10.1016/j.physletb.2015.06.055}{{\em Phys. Lett. B}
  {\bfseries 748} (2015) 5--8}
  [\href{http://arxiv.org/abs/1505.03893}{{\ttfamily arXiv:1505.03893}}]
  [\href{http://inspirehep.net/search?p=find+Boughezal:2015aha}{{\ttfamily
  InSPIRE}}].

\bibitem{Chen:2016zka}
X.~Chen, J.~Cruz-Martinez, T.~Gehrmann, E.~W.~N. Glover, and M.~Jaquier, {\it
  {NNLO QCD corrections to Higgs boson production at large transverse
  momentum}},  \href{http://dx.doi.org/10.1007/JHEP10(2016)066}{{\em JHEP}
  {\bfseries 10} (2016) 066} [\href{http://arxiv.org/abs/1607.08817}{{\ttfamily
  arXiv:1607.08817}}]
  [\href{http://inspirehep.net/search?p=find+Chen:2016zka}{{\ttfamily
  InSPIRE}}].

\bibitem{Henn:2023vbd}
J.~M. Henn, J.~Lim, and W.~J. Torres~Bobadilla, {\it {First look at the
  evaluation of three-loop non-planar Feynman diagrams for Higgs plus jet
  production}},  \href{http://dx.doi.org/10.1007/JHEP05(2023)026}{{\em JHEP}
  {\bfseries 05} (2023) 026} [\href{http://arxiv.org/abs/2302.12776}{{\ttfamily
  arXiv:2302.12776}}]
  [\href{http://inspirehep.net/search?p=find+Henn:2023vbd}{{\ttfamily
  InSPIRE}}].

\bibitem{Gehrmann:2023etk}
T.~Gehrmann, P.~Jakub{\v{c}}{\'\i}k, C.~C. Mella, N.~Syrrakos, and L.~Tancredi,
  {\it {Two-loop helicity amplitudes for $H+$jet production to higher orders in
  the dimensional regulator}},
  \href{http://dx.doi.org/10.1007/JHEP04(2023)016}{{\em JHEP} {\bfseries 04}
  (2023) 016} [\href{http://arxiv.org/abs/2301.10849}{{\ttfamily
  arXiv:2301.10849}}]
  [\href{http://inspirehep.net/search?p=find+Gehrmann:2023etk}{{\ttfamily
  InSPIRE}}].

\bibitem{Gehrmann:2024tds}
T.~Gehrmann, J.~Henn, P.~Jakub{\v{c}}{\'\i}k, J.~Lim, C.~C. Mella, N.~Syrrakos,
  L.~Tancredi, and W.~J. Torres~Bobadilla, {\it {Graded transcendental
  functions: an application to four-point amplitudes with one off-shell leg}},
  \href{http://dx.doi.org/10.1007/JHEP12(2024)215}{{\em JHEP} {\bfseries 12}
  (2024) 215} [\href{http://arxiv.org/abs/2410.19088}{{\ttfamily
  arXiv:2410.19088}}]
  [\href{http://inspirehep.net/search?p=find+Gehrmann:2024tds}{{\ttfamily
  InSPIRE}}].

\bibitem{Ciccolini:2007jr}
M.~Ciccolini, A.~Denner, and S.~Dittmaier, {\it {Strong and electroweak
  corrections to the production of Higgs + 2jets via weak interactions at the
  LHC}},  \href{http://dx.doi.org/10.1103/PhysRevLett.99.161803}{{\em Phys.
  Rev. Lett.} {\bfseries 99} (2007) 161803}
  [\href{http://arxiv.org/abs/0707.0381}{{\ttfamily arXiv:0707.0381}}]
  [\href{http://inspirehep.net/search?p=find+Ciccolini:2007jr}{{\ttfamily
  InSPIRE}}].

\bibitem{Denner:2014cla}
A.~Denner, S.~Dittmaier, S.~Kallweit, and A.~M{\"u}ck, {\it {HAWK 2.0: A Monte
  Carlo program for Higgs production in vector-boson fusion and Higgs strahlung
  at hadron colliders}},
  \href{http://dx.doi.org/10.1016/j.cpc.2015.04.021}{{\em Comput. Phys.
  Commun.} {\bfseries 195} (2015) 161--171}
  [\href{http://arxiv.org/abs/1412.5390}{{\ttfamily arXiv:1412.5390}}]
  [\href{http://inspirehep.net/search?p=find+Denner:2014cla}{{\ttfamily
  InSPIRE}}].

\bibitem{Ciccolini:2003jy}
M.~L. Ciccolini, S.~Dittmaier, and M.~Kramer, {\it {Electroweak radiative
  corrections to associated WH and ZH production at hadron colliders}},
  \href{http://dx.doi.org/10.1103/PhysRevD.68.073003}{{\em Phys. Rev. D}
  {\bfseries 68} (2003) 073003}
  [\href{http://arxiv.org/abs/hep-ph/0306234}{{\ttfamily hep-ph/0306234}}]
  [\href{http://inspirehep.net/search?p=find+Ciccolini:2003jy}{{\ttfamily
  InSPIRE}}].

\bibitem{Denner:2011id}
A.~Denner, S.~Dittmaier, S.~Kallweit, and A.~Muck, {\it {Electroweak
  corrections to Higgs-strahlung off W/Z bosons at the Tevatron and the LHC
  with HAWK}},  \href{http://dx.doi.org/10.1007/JHEP03(2012)075}{{\em JHEP}
  {\bfseries 03} (2012) 075} [\href{http://arxiv.org/abs/1112.5142}{{\ttfamily
  arXiv:1112.5142}}]
  [\href{http://inspirehep.net/search?p=find+Denner:2011id}{{\ttfamily
  InSPIRE}}].

\bibitem{Obul:2018psx}
P.~Obul, S.~Dulat, T.-J. Hou, A.~Tursun, and N.~Yalkun, {\it {Next-to-leading
  order QCD and electroweak corrections to Higgs-strahlung processes at the
  LHC}},  \href{http://dx.doi.org/10.1088/1674-1137/42/9/093105}{{\em Chin.
  Phys. C} {\bfseries 42} (2018) 093105}
  [\href{http://arxiv.org/abs/1801.06851}{{\ttfamily arXiv:1801.06851}}]
  [\href{http://inspirehep.net/search?p=find+Obul:2018psx}{{\ttfamily
  InSPIRE}}].

\bibitem{Granata:2017iod}
F.~Granata, J.~M. Lindert, C.~Oleari, and S.~Pozzorini, {\it {NLO QCD+EW
  predictions for HV and HV +jet production including parton-shower effects}},
  \href{http://dx.doi.org/10.1007/JHEP09(2017)012}{{\em JHEP} {\bfseries 09}
  (2017) 012} [\href{http://arxiv.org/abs/1706.03522}{{\ttfamily
  arXiv:1706.03522}}]
  [\href{http://inspirehep.net/search?p=find+Granata:2017iod}{{\ttfamily
  InSPIRE}}].

\bibitem{Zhang:2014gcy}
Y.~Zhang, W.-G. Ma, R.-Y. Zhang, C.~Chen, and L.~Guo, {\it {QCD NLO and EW NLO
  corrections to $t\bar{t}H$ production with top quark decays at hadron
  collider}},  \href{http://dx.doi.org/10.1016/j.physletb.2014.09.022}{{\em
  Phys. Lett. B} {\bfseries 738} (2014) 1--5}
  [\href{http://arxiv.org/abs/1407.1110}{{\ttfamily arXiv:1407.1110}}]
  [\href{http://inspirehep.net/search?p=find+Zhang:2014gcy}{{\ttfamily
  InSPIRE}}].

\bibitem{Frixione:2014qaa}
S.~Frixione, V.~Hirschi, D.~Pagani, H.~S. Shao, and M.~Zaro, {\it {Weak
  corrections to Higgs hadroproduction in association with a top-quark pair}},
  \href{http://dx.doi.org/10.1007/JHEP09(2014)065}{{\em JHEP} {\bfseries 09}
  (2014) 065} [\href{http://arxiv.org/abs/1407.0823}{{\ttfamily
  arXiv:1407.0823}}]
  [\href{http://inspirehep.net/search?p=find+Frixione:2014qaa}{{\ttfamily
  InSPIRE}}].

\bibitem{Denner:2016wet}
A.~Denner, J.-N. Lang, M.~Pellen, and S.~Uccirati, {\it {Higgs production in
  association with off-shell top-antitop pairs at NLO EW and QCD at the LHC}},
  \href{http://dx.doi.org/10.1007/JHEP02(2017)053}{{\em JHEP} {\bfseries 02}
  (2017) 053} [\href{http://arxiv.org/abs/1612.07138}{{\ttfamily
  arXiv:1612.07138}}]
  [\href{http://inspirehep.net/search?p=find+Denner:2016wet}{{\ttfamily
  InSPIRE}}].

\bibitem{Bonetti:2020hqh}
M.~Bonetti, E.~Panzer, V.~A. Smirnov, and L.~Tancredi, {\it {Two-loop mixed
  QCD-EW corrections to $gg \to Hg$}},
  \href{http://dx.doi.org/10.1007/JHEP11(2020)045}{{\em JHEP} {\bfseries 11}
  (2020) 045} [\href{http://arxiv.org/abs/2007.09813}{{\ttfamily
  arXiv:2007.09813}}]
  [\href{http://inspirehep.net/search?p=find+Bonetti:2020hqh}{{\ttfamily
  InSPIRE}}].

\bibitem{Becchetti:2020wof}
M.~Becchetti, R.~Bonciani, V.~Del~Duca, V.~Hirschi, F.~Moriello, and
  A.~Schweitzer, {\it {Next-to-leading order corrections to light-quark mixed
  QCD-EW contributions to Higgs boson production}},
  \href{http://dx.doi.org/10.1103/PhysRevD.103.054037}{{\em Phys. Rev. D}
  {\bfseries 103} (2021) 054037}
  [\href{http://arxiv.org/abs/2010.09451}{{\ttfamily arXiv:2010.09451}}]
  [\href{http://inspirehep.net/search?p=find+Becchetti:2020wof}{{\ttfamily
  InSPIRE}}].

\bibitem{Davies:2023npk}
J.~Davies, K.~Sch\"onwald, M.~Steinhauser, and H.~Zhang, {\it {Next-to-leading
  order electroweak corrections to gg \textrightarrow{} HH and gg
  \textrightarrow{} gH in the large-m$_{t}$ limit}},
  \href{http://dx.doi.org/10.1007/JHEP10(2023)033}{{\em JHEP} {\bfseries 10}
  (2023) 033} [\href{http://arxiv.org/abs/2308.01355}{{\ttfamily
  arXiv:2308.01355}}]
  [\href{http://inspirehep.net/search?p=find+Davies:2023npk}{{\ttfamily
  InSPIRE}}].

\bibitem{Gao:2023bll}
J.~Gao, X.-M. Shen, G.~Wang, L.~L. Yang, and B.~Zhou, {\it {Probing the Higgs
  boson trilinear self-coupling through Higgs boson+jet production}},
  \href{http://dx.doi.org/10.1103/PhysRevD.107.115017}{{\em Phys. Rev. D}
  {\bfseries 107} (2023) 115017}
  [\href{http://arxiv.org/abs/2302.04160}{{\ttfamily arXiv:2302.04160}}]
  [\href{http://inspirehep.net/search?p=find+Gao:2023bll}{{\ttfamily
  InSPIRE}}].

\bibitem{Haisch:2024nzv}
U.~Haisch and M.~Niggetiedt, {\it {Exact two-loop amplitudes for Higgs plus jet
  production with a cubic Higgs self-coupling}},
  \href{http://dx.doi.org/10.1007/JHEP10(2024)236}{{\em JHEP} {\bfseries 10}
  (2024) 236} [\href{http://arxiv.org/abs/2408.13186}{{\ttfamily
  arXiv:2408.13186}}]
  [\href{http://inspirehep.net/search?p=find+Haisch:2024nzv}{{\ttfamily
  InSPIRE}}].

\bibitem{Kotikov:1990kg}
A.~V. Kotikov, {\it {Differential equations method: New technique for massive
  Feynman diagrams calculation}},
  \href{http://dx.doi.org/10.1016/0370-2693(91)90413-K}{{\em Phys. Lett. B}
  {\bfseries 254} (1991) 158--164}
  [\href{http://inspirehep.net/search?p=find+Kotikov:1990kg}{{\ttfamily
  InSPIRE}}].

\bibitem{Remiddi:1997ny}
E.~Remiddi, {\it {Differential equations for Feynman graph amplitudes}},
  \href{http://dx.doi.org/10.1007/BF03185566}{{\em Nuovo Cim. A} {\bfseries
  110} (1997) 1435--1452}
  [\href{http://arxiv.org/abs/hep-th/9711188}{{\ttfamily hep-th/9711188}}]
  [\href{http://inspirehep.net/search?p=find+Remiddi:1997ny}{{\ttfamily
  InSPIRE}}].

\bibitem{Caffo:2008aw}
M.~Caffo, H.~Czyz, M.~Gunia, and E.~Remiddi, {\it {BOKASUN: A Fast and precise
  numerical program to calculate the Master Integrals of the two-loop sunrise
  diagrams}},  \href{http://dx.doi.org/10.1016/j.cpc.2008.10.011}{{\em Comput.
  Phys. Commun.} {\bfseries 180} (2009) 427--430}
  [\href{http://arxiv.org/abs/0807.1959}{{\ttfamily arXiv:0807.1959}}]
  [\href{http://inspirehep.net/search?p=find+Caffo:2008aw}{{\ttfamily
  InSPIRE}}].

\bibitem{Czakon:2008zk}
M.~Czakon, {\it {Tops from Light Quarks: Full Mass Dependence at Two-Loops in
  QCD}},  \href{http://dx.doi.org/10.1016/j.physletb.2008.05.028}{{\em Phys.
  Lett. B} {\bfseries 664} (2008) 307--314}
  [\href{http://arxiv.org/abs/0803.1400}{{\ttfamily arXiv:0803.1400}}]
  [\href{http://inspirehep.net/search?p=find+Czakon:2008zk}{{\ttfamily
  InSPIRE}}].

\bibitem{Liu:2017jxz}
X.~Liu, Y.-Q. Ma, and C.-Y. Wang, {\it {A Systematic and Efficient Method to
  Compute Multi-loop Master Integrals}},
  \href{http://dx.doi.org/10.1016/j.physletb.2018.02.026}{{\em Phys. Lett. B}
  {\bfseries 779} (2018) 353--357}
  [\href{http://arxiv.org/abs/1711.09572}{{\ttfamily arXiv:1711.09572}}]
  [\href{http://inspirehep.net/search?p=find+Liu:2017jxz}{{\ttfamily
  InSPIRE}}].

\bibitem{Liu:2021wks}
X.~Liu and Y.-Q. Ma, {\it {Multiloop corrections for collider processes using
  auxiliary mass flow}},
  \href{http://dx.doi.org/10.1103/PhysRevD.105.L051503}{{\em Phys. Rev. D}
  {\bfseries 105} (2022) L051503}
  [\href{http://arxiv.org/abs/2107.01864}{{\ttfamily arXiv:2107.01864}}]
  [\href{http://inspirehep.net/search?p=find+Liu:2021wks}{{\ttfamily
  InSPIRE}}].

\bibitem{Liu:2022mfb}
Z.-F. Liu and Y.-Q. Ma, {\it {Determining Feynman Integrals with Only Input
  from Linear Algebra}},
  \href{http://dx.doi.org/10.1103/PhysRevLett.129.222001}{{\em Phys. Rev.
  Lett.} {\bfseries 129} (2022) 222001}
  [\href{http://arxiv.org/abs/2201.11637}{{\ttfamily arXiv:2201.11637}}]
  [\href{http://inspirehep.net/search?p=find+Liu:2022mfb}{{\ttfamily
  InSPIRE}}].

\bibitem{Liu:2022chg}
X.~Liu and Y.-Q. Ma, {\it {AMFlow: A Mathematica package for Feynman integrals
  computation via auxiliary mass flow}},
  \href{http://dx.doi.org/10.1016/j.cpc.2022.108565}{{\em Comput. Phys.
  Commun.} {\bfseries 283} (2023) 108565}
  [\href{http://arxiv.org/abs/2201.11669}{{\ttfamily arXiv:2201.11669}}]
  [\href{http://inspirehep.net/search?p=find+Liu:2022chg}{{\ttfamily
  InSPIRE}}].

\bibitem{Hahn:2000kx}
T.~Hahn, {\it {Generating Feynman diagrams and amplitudes with FeynArts 3}},
  \href{http://dx.doi.org/10.1016/S0010-4655(01)00290-9}{{\em Comput. Phys.
  Commun.} {\bfseries 140} (2001) 418--431}
  [\href{http://arxiv.org/abs/hep-ph/0012260}{{\ttfamily hep-ph/0012260}}]
  [\href{http://inspirehep.net/search?p=find+Hahn:2000kx}{{\ttfamily
  InSPIRE}}].

\bibitem{Melnikov:2016qoc}
K.~Melnikov, L.~Tancredi, and C.~Wever, {\it {Two-loop $gg \to Hg$ amplitude
  mediated by a nearly massless quark}},
  \href{http://dx.doi.org/10.1007/JHEP11(2016)104}{{\em JHEP} {\bfseries 11}
  (2016) 104} [\href{http://arxiv.org/abs/1610.03747}{{\ttfamily
  arXiv:1610.03747}}]
  [\href{http://inspirehep.net/search?p=find+Melnikov:2016qoc}{{\ttfamily
  InSPIRE}}].

\bibitem{Bi:2023bnq}
H.-Y. Bi, L.-H. Huang, R.-J. Huang, Y.-Q. Ma, and H.-M. Yu, {\it {Electroweak
  Corrections to Double Higgs Production at the LHC}},
  \href{http://dx.doi.org/10.22323/1.478.0120}{{\em Phys. Rev. Lett.}
  {\bfseries 132} (2024) 231802}
  [\href{http://arxiv.org/abs/2311.16963}{{\ttfamily arXiv:2311.16963}}]
  [\href{http://inspirehep.net/search?p=find+Bi:2023bnq}{{\ttfamily InSPIRE}}].

\bibitem{calcloop}
 \url{https://e.gitee.com/multiloop-pku/repos/multiloop-pku/calcloop/sources}.

\bibitem{Guan:2024byi}
X.~Guan, X.~Liu, Y.-Q. Ma, and W.-H. Wu, {\it {Blade: A package for
  block-triangular form improved Feynman integrals decomposition}},
  \href{http://dx.doi.org/10.1016/j.cpc.2025.109538}{{\em Comput. Phys.
  Commun.} {\bfseries 310} (2025) 109538}
  [\href{http://arxiv.org/abs/2405.14621}{{\ttfamily arXiv:2405.14621}}]
  [\href{http://inspirehep.net/search?p=find+Guan:2024byi}{{\ttfamily
  InSPIRE}}].

\bibitem{Peraro:2019svx}
T.~Peraro, {\it {$\text{FiniteFlow}$: multivariate functional reconstruction
  using finite fields and dataflow graphs}},
  \href{http://dx.doi.org/10.1007/JHEP07(2019)031}{{\em JHEP} {\bfseries 07}
  (2019) 031} [\href{http://arxiv.org/abs/1905.08019}{{\ttfamily
  arXiv:1905.08019}}]
  [\href{http://inspirehep.net/search?p=find+Peraro:2019svx}{{\ttfamily
  InSPIRE}}].

\bibitem{vanHameren:2007pt}
A.~van Hameren, {\it {PARNI for importance sampling and density estimation}},
  {\em Acta Phys. Polon. B} {\bfseries 40} (2009) 259--272
  [\href{http://arxiv.org/abs/0710.2448}{{\ttfamily arXiv:0710.2448}}]
  [\href{http://inspirehep.net/search?p=find+vanHameren:2007pt}{{\ttfamily
  InSPIRE}}].

\bibitem{Alwall:2014hca}
J.~Alwall, R.~Frederix, S.~Frixione, V.~Hirschi, F.~Maltoni, O.~Mattelaer,
  H.~S. Shao, T.~Stelzer, P.~Torrielli, and M.~Zaro, {\it {The automated
  computation of tree-level and next-to-leading order differential cross
  sections, and their matching to parton shower simulations}},
  \href{http://dx.doi.org/10.1007/JHEP07(2014)079}{{\em JHEP} {\bfseries 07}
  (2014) 079} [\href{http://arxiv.org/abs/1405.0301}{{\ttfamily
  arXiv:1405.0301}}]
  [\href{http://inspirehep.net/search?p=find+Alwall:2014hca}{{\ttfamily
  InSPIRE}}].

\bibitem{Huang:2024qan}
R.-J. Huang, D.-S. Jian, Y.-Q. Ma, D.-M. Mu, and W.-H. Wu, {\it {Efficient
  computation of one-loop Feynman integrals and fixed-branch integrals to high
  orders in {\ensuremath{\varepsilon}}}},
  \href{http://dx.doi.org/10.1103/PhysRevD.111.094028}{{\em Phys. Rev. D}
  {\bfseries 111} (2025) 094028}
  [\href{http://arxiv.org/abs/2412.21054}{{\ttfamily arXiv:2412.21054}}]
  [\href{http://inspirehep.net/search?p=find+Huang:2024qan}{{\ttfamily
  InSPIRE}}].

\bibitem{Alwall:2006yp}
J.~Alwall {\em et al.}, {\it {A Standard format for Les Houches event files}},
  \href{http://dx.doi.org/10.1016/j.cpc.2006.11.010}{{\em Comput. Phys.
  Commun.} {\bfseries 176} (2007) 300--304}
  [\href{http://arxiv.org/abs/hep-ph/0609017}{{\ttfamily hep-ph/0609017}}]
  [\href{http://inspirehep.net/search?p=find+Alwall:2006yp}{{\ttfamily
  InSPIRE}}].

\bibitem{ParticleDataGroup:2022pth}
{\bfseries Particle Data Group} , R.~L. Workman {\em et al.}, {\it {Review of
  Particle Physics}},  \href{http://dx.doi.org/10.1093/ptep/ptac097}{{\em PTEP}
  {\bfseries 2022} (2022) 083C01}
  [\href{http://inspirehep.net/search?p=find+ParticleDataGroup:2022pth}{{\ttfamily
  InSPIRE}}].

\bibitem{Denner:2019vbn}
A.~Denner and S.~Dittmaier, {\it {Electroweak Radiative Corrections for
  Collider Physics}},
  \href{http://dx.doi.org/10.1016/j.physrep.2020.04.001}{{\em Phys. Rept.}
  {\bfseries 864} (2020) 1--163}
  [\href{http://arxiv.org/abs/1912.06823}{{\ttfamily arXiv:1912.06823}}]
  [\href{http://inspirehep.net/search?p=find+Denner:2019vbn}{{\ttfamily
  InSPIRE}}].

\bibitem{Denner:2014bna}
A.~Denner, S.~Dittmaier, M.~Hecht, and C.~Pasold, {\it {NLO QCD and electroweak
  corrections to W+{\textbackslash}gamma{\textbackslash} production with
  leptonic W-boson decays}},
  \href{http://dx.doi.org/10.1007/JHEP04(2015)018}{{\em JHEP} {\bfseries 04}
  (2015) 018} [\href{http://arxiv.org/abs/1412.7421}{{\ttfamily
  arXiv:1412.7421}}]
  [\href{http://inspirehep.net/search?p=find+Denner:2014bna}{{\ttfamily
  InSPIRE}}].

\bibitem{NNPDF:2017mvq}
{\bfseries NNPDF} , R.~D. Ball {\em et al.}, {\it {Parton distributions from
  high-precision collider data}},
  \href{http://dx.doi.org/10.1140/epjc/s10052-017-5199-5}{{\em Eur. Phys. J. C}
  {\bfseries 77} (2017) 663} [\href{http://arxiv.org/abs/1706.00428}{{\ttfamily
  arXiv:1706.00428}}]
  [\href{http://inspirehep.net/search?p=find+NNPDF:2017mvq}{{\ttfamily
  InSPIRE}}].

\bibitem{Buckley:2014ana}
A.~Buckley, J.~Ferrando, S.~Lloyd, K.~Nordstr\"om, B.~Page, M.~R\"ufenacht,
  M.~Sch\"onherr, and G.~Watt, {\it {LHAPDF6: parton density access in the LHC
  precision era}},
  \href{http://dx.doi.org/10.1140/epjc/s10052-015-3318-8}{{\em Eur. Phys. J. C}
  {\bfseries 75} (2015) 132} [\href{http://arxiv.org/abs/1412.7420}{{\ttfamily
  arXiv:1412.7420}}]
  [\href{http://inspirehep.net/search?p=find+Buckley:2014ana}{{\ttfamily
  InSPIRE}}].

\bibitem{Harlander:2020cyh}
R.~V. Harlander, S.~Y. Klein, and M.~Lipp, {\it {FeynGame}},
  \href{http://dx.doi.org/10.1016/j.cpc.2020.107465}{{\em Comput. Phys.
  Commun.} {\bfseries 256} (2020) 107465}
  [\href{http://arxiv.org/abs/2003.00896}{{\ttfamily arXiv:2003.00896}}]
  [\href{http://inspirehep.net/search?p=find+Harlander:2020cyh}{{\ttfamily
  InSPIRE}}].

\bibitem{Chen:2025sgo}
L.-B. Chen, H.~T. Li, and W.-L. Sang, {\it {Electroweak corrections to
  Higgs+jet production in gluon fusion}},
  [\href{http://arxiv.org/abs/2507.20882}{{\ttfamily arXiv:2507.20882}}]
  [\href{http://inspirehep.net/search?p=find+Chen:2025sgo}{{\ttfamily
  InSPIRE}}].

\end{thebibliography}
